\documentclass[twocolumn,showpacs,preprintnumbers,amsmath,amssymb,prl,superscriptaddress]{revtex4}
\usepackage{graphicx}% Include figure files
\usepackage{dcolumn}% Align table columns on decimal point
\usepackage{bm}% bold math
\usepackage{amsmath}
\usepackage{amssymb}

\newcommand{\CQCTsyd}{Centre for Quantum Computer Technology, School of Electrical Engineering \& Telecommunications,
\\ University of New South Wales, Sydney NSW 2052, Australia.}
\newcommand{\CQCTmel}{Centre for Quantum Computer Technology, School of Physics,
University of Melbourne, Melbourne VIC 3010, Australia.}
\newcommand{\TKKa}{Department of Applied Physics/COMP, Helsinki University of
  Technology, P.O.~Box 5100, FI-02015 TKK, Finland.}
\newcommand{\TKKb}{Low Temperature Laboratory, Helsinki University of Technology,
  P.O.~Box 3500, FI-02015 TKK, Finland.}

\begin{document}

%\nofiles
\title{Probe and Control of the Reservoir Density of States in Single-Electron Devices}

%\preprint{}

\author{M.~M\"{o}tt\"{o}nen}
\address{{\rm These authors contributed equally.}}
\affiliation{\CQCTsyd} \affiliation{\TKKa} \affiliation{\TKKb}

\author{K.~Y.~Tan}
\address{{\rm These authors contributed equally.}}
\affiliation{\CQCTsyd}

\author{K.~W.~Chan}
\affiliation{\CQCTsyd}

\author{F.~A.~Zwanenburg}
\affiliation{\CQCTsyd}

\author{W.~H.~Lim}
\affiliation{\CQCTsyd}

\author{C.~C.~Escott}
\affiliation{\CQCTsyd}

\author{J.-M.~Pirkkalainen}
\affiliation{\CQCTsyd} \affiliation{\TKKa}

\author{A.~Morello}
\affiliation{\CQCTsyd}

\author{C.~Yang}
\affiliation{\CQCTmel}

\author{J.~A.~van Donkelaar}
\affiliation{\CQCTmel}

\author{A.~D.~C.~Alves}
\affiliation{\CQCTmel}

\author{D.~N.~Jamieson}
\affiliation{\CQCTmel}

\author{L.~C.~L.~Hollenberg}
\affiliation{\CQCTmel}

\author{A.~S.~Dzurak}
\affiliation{\CQCTsyd}

\date{\today}

\begin{abstract}
We present a systematic study of quasi-one-dimensional density of
states (DOS) in electron accumulation layers near a Si--SiO$_2$
interface. In the experiments we have employed two conceptually
different objects to probe DOS, namely, a phosphorus donor and a
quantum dot, both operating in the single-electron tunneling regime.
We demonstrate how the peaks in DOS can be moved in the transport
window independently of the other device properties, and in
agreement with the theoretical analysis. This method introduces a
fast and convenient way of identifying excited states in these
emerging nanostructures.
\end{abstract}


\maketitle

The appearance of discrete energy spectra is often regarded as a
fundamental property of quantum systems. However, the limit of
vanishing energy spacing can be met in many mesoscopic systems at
the interface of the quantum and classical regimes meaning that the
energy levels can be treated as a continuum. This gives rise to the
concept of density of states (DOS), the integral of which over an
energy domain yields the number of quantum states in that region.
The DOS concept has been successfully applied in explaining
transport, absorption and emission, and quantum statistical
phenomena in such devices~\cite{AverinLikharev}. In particular, DOS
is a key element in the low-temperature behavior of
metal-oxide-semiconductor field-effect transistors~\cite{AndoRMP}
(MOSFETs) which constitute the cornerstone of information processing
circuits today. Since device miniaturization has now reached the
point where quantum effects of single atoms can dominate the
operation
characteristics~\cite{SellierPRL06,CalvetPRL07,LansbergenNatPhys08,Tan2009},
there is an urgent need to understand and distinguish continuum DOS
effects from the discrete quantum behavior. In this paper we present
a systematic study of the reservoir DOS in two gated MOSFET
nanostructures with quantum channels defined by either a quantum dot
or the extreme case of a single donor atom.

\begin{figure}[th!] \center
\includegraphics[width=0.99\linewidth]{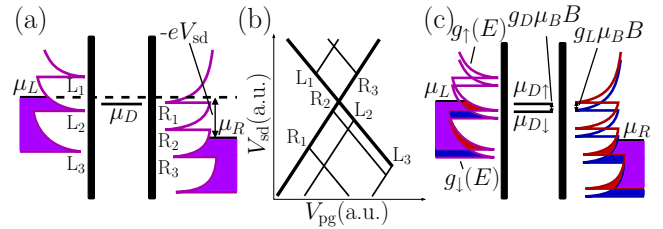}
\caption{ \label{fig1} (Color online) (a) Schematics of single-electron tunneling through a discrete quantum level, $\mu_D$, together with the reservoir Fermi levels, $\mu_{L/R}$, and density of states. (b) Schematic stability diagram showing conductance through the discrete quantum level as a function of the source--drain bias and the plunger gate voltage. The numbered lines of increased conductance arise from the corresponding DOS peaks in panel (a). (c) Zeeman splitting of the reservoir and discrete energy levels in a magnetic field.}% end caption
\end{figure}

In a silicon MOSFET a positive gate voltage is applied to induce an
electron layer directly below the Si--SiO$_2$ interface. Hence, the
electron dynamics is essentially limited to two dimensions, leading
ideally to a constant DOS. In a narrow channel, however, the
continuum approximation is only valid in the longest direction,
giving rise to quasi-one-dimensional (Q1D) DOS with highly
non-uniform characteristics, see Fig.~\ref{fig1}(a). Conductance
modulations attributed to Q1D density of states were first observed
in a parallel configuration of 250 narrow MOSFET
channels~\cite{1st1dDOSwarren} and later in a single
channel~\cite{1dDOSmorimoto,1dDOSmatsuoka,1dDOStakeuchi}. Since the
cumulative conductance through all occupied states was measured in
these experiments, the studies were limited to rather low electron
densities in the channel. The Q1D density of states is also
intimately related to ballistic electron transport through quantum
point contacts~\cite{vanWees1988,Wharam1988}, and has been probed in
single carbon nanotubes using scanning tunneling
microscopy~\cite{Wildoer1998,Venema2000}.

In single-electron transport through discrete quantum
states~\cite{AverinLikharev}, the current is directly proportional
to the reservoir DOS at a given energy, see Fig.~\ref{fig1}.
Considerable effort has been directed towards the study of local
density of states of a reservoir in the vicinity of an impurity atom
in a GaAs quantum well~\cite{Holder1999,Schmidt1997,Schmidt1996}.
Here, the local DOS was dominated by disorder due to impurity
scattering, resulting in reproducible but irregular features in the
conductance which behaved in a complicated way as a function of
magnetic field. Thus the term local DOS fluctuations was introduced.
In our case, the behavior is less complicated, and hence we do not
refer to peaks in DOS as fluctuations.

In gated devices, local DOS effects have also been observed in
tunneling through a vertical quantum dot~\cite{Kouwenhoven1997}, for
which a schematic stability diagram is shown in Fig.~\ref{fig1}(b).
Recently, gated quantum dots have attracted great interest due to
their tunability, and lines in the stability diagrams not attributed
to excited states have been observed in various
structures~\cite{Bjork2004,Zwanenburg2009,Leturcq2009}. In
Refs.~\cite{Zwanenburg2009,Leturcq2009} however, these lines were
due to phonon modes, whereas in Ref.~\cite{Bjork2004} they were
suggested to arise from reservoir DOS but were not studied
thoroughly.

\begin{figure}[th!] \center
\includegraphics[width=0.95\linewidth]{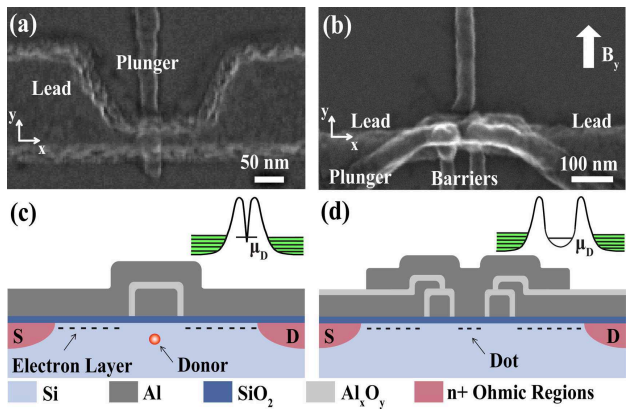}
\caption{ \label{fig2} (Color online) (a) Top view of Structure A and (b) Structure B and a schematic cross section of (c) Structure A and (d) Structure B in the $y=0$ plane. The insets show schematic potential landscapes along $x$.}% end caption
\end{figure}

Motivated by the Kane proposal~\cite{KaneNature98} for a quantum
computer based on shallow donors in silicon~\cite{Vrijen2000}, there
has been considerable development in electron transport through
gated single
donors~\cite{SellierPRL06,CalvetPRL07,LansbergenNatPhys08,Tan2009}.
In these experiments, conductance lines attributed to reservoir DOS
have been observed but, again, not studied in detail. Indeed, it has
not yet been proven that these features are due to the reservoir
DOS. In this paper we present a systematic study of the reservoir
DOS in two different nanostructures, namely, in a recently
introduced double-gated single-donor transistor~\cite{Tan2009}
(Structure A) and in a novel multi-gated silicon quantum
dot~\cite{WeeHan} (Structure B), shown in Fig.~\ref{fig2}. In
contrast to previous studies of DOS, the double- and multi-gated
designs allow us to map conveniently the reservoir DOS at a wide
range of energies and electron densities.

The high-purity [001] silicon substrates for both device structures
went through similar fabrication steps to those described in detail
in Ref.~\cite{Tan2009}. In Structure B, however, there are three
layers of metallic gates deposited with no implanted donors
underneath. The working principle of both devices is that an
accumulation layer of electrons is induced directly below the
Si--SiO$_2$ interface using the lead gate voltage $V_\textrm{lg}$.
This layer constitutes the reservoirs for single-electron tunneling
through a single donor or a quantum dot, the electrochemical
potentials of which can be tuned by the plunger gate voltage
$V_\textrm{pg}$. For Structure A, the plunger gate works also as a
barrier gate, depleting the reservoirs in the vicinity of the donor.
For Structure B, however, we have separate barrier gates which can
be used to tune the coupling of the quantum dot to the left and
right reservoir independently. The electron densities in the
reservoirs can also be controlled independently, but for simplicity,
we keep them at the same value here. As the plunger gate voltage is
increased, the electrochemical potential of the donor or dot,
$\mu_D$, shifts down, eventually entering the source--drain bias
window leading to single-electron tunneling through the device, see
Fig.~\ref{fig1}(a--b).

The sequential tunneling rate is directly proportional to the
reservoir density of states, which leads to a peak in the
source--drain current if $\mu_D$ is aligned with a peak in the
density of states. Critically in our devices, we can shift the Fermi
levels of the reservoirs with respect to the conduction band minima
by changing the lead gate voltage, $V_\textrm{lg}$, which in turn
moves the DOS peaks with respect to the transport window. The effect
of the lead gate voltage on $\mu_D$ can be compensated by the
plunger gate~\cite{FN1x}. Therefore, if we observe a conductance
peak to shift with $V_\textrm{lg}$ in the stability diagram, we can
identify it to arise from a peak in the reservoir DOS, clearly
distinct from features due to excited states in the dot or donor
which do not move with $V_\textrm{lg}$. Previously, DOS peaks have
been probed by changing the temperature~\cite{Schmidt1996} or
magnetic field~\cite{Jouault2009}, both of which also have an impact
on the donor or the dot, and which are many orders of magnitude
slower than the method presented here.

The measurements were carried out in $^3$He--$^4$He dilution
refrigerators below 100~mK temperatures. The conductance was
measured using standard lock-in techniques and the direct current
was measured from the same signal after low-pass filtering the
modulation arising from the 10--50~$\mu$V lock-in excitation. For
Structure B, only the direct current was recorded and the
differential conductance was extracted numerically.

\begin{figure}[th!] \center
\includegraphics[width=0.9\linewidth]{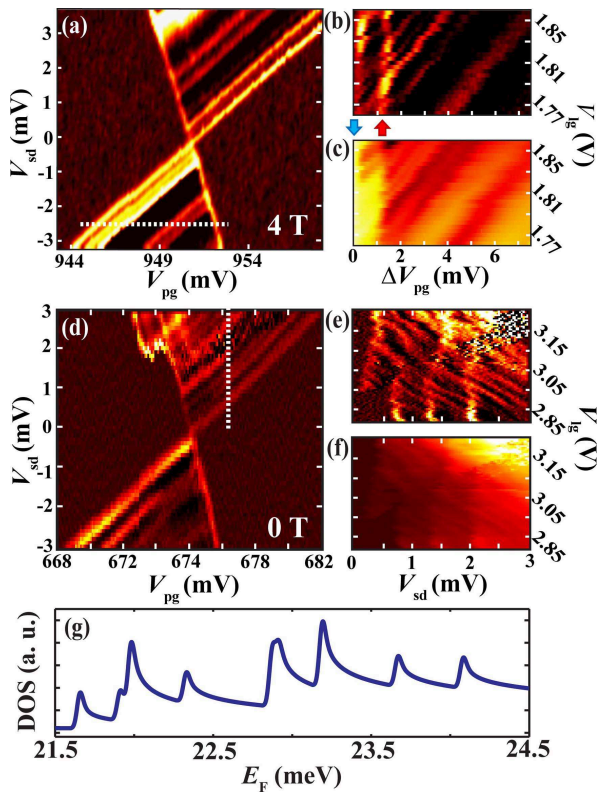}
\caption{ \label{fig3} (Color online) Conductance of (a) Structure A
with $V_\textrm{lg}=1.87$~V and $B_\parallel= 4$~T, and (d)
Structure B with $V_\textrm{lg}=3.00$~V, $V_\textrm{lb}=0.52$~V,
$V_\textrm{rb}=0.68$~V, and
$B_\parallel= 0$~T as a function of the source--drain bias and the plunger gate voltage. The dashed line in panel (a) represents the trace shown in panels (b) and (c) for $V_\textrm{sd}=-2.5$~mV as a function of the lead gate voltage for differential conductance and direct current, respectively. Panels (e) and (f) correspond to the dashed trace shown in panel (d) for $\Delta V_\textrm{pg}=2.1$~mV from the transport window at $V_\textrm{sd}=0$. The capacitive coupling of the lead gate to the donor or the dot has been compensated by the plunger gate in the traces. Panel (g) shows the theoretical prediction for DOS corresponding to panels (b,c). Here, Eq.~\eqref{eq1} has been convolved with the Gaussian $\exp\{-E_F^2/[2(500~\mu\textrm{eV})^2]\}$ to account for broadening due to experimental nonidealities.}% end caption
\end{figure}

Figure~\ref{fig3} shows measured stability diagrams for Structure A
[panel (a)] and B [panel (d)]. In Structure A the coupling of the
donor to the left reservoir is much weaker than to the right
reservoir, and hence only lines with positive slopes corresponding
to the left reservoir are visible. This is justified by the
sequential tunneling model, in which the current through the device
is given by
$I=2e\Gamma_\textrm{in}\Gamma_\textrm{out}/(2\Gamma_\textrm{in}+\Gamma_\textrm{out})$,
where $\Gamma_\textrm{in/out}$ is the in/out-tunneling rate. Thus
the current is determined by the small rate, independent of whether
it corresponds to in- or out-tunneling. The tunneling rates in
Structure B can be tuned to be symmetric, but considerable asymmetry
persists in the working point of Fig.~\ref{fig3}.

\begin{figure}[th!] \center
\includegraphics[width=0.9\linewidth]{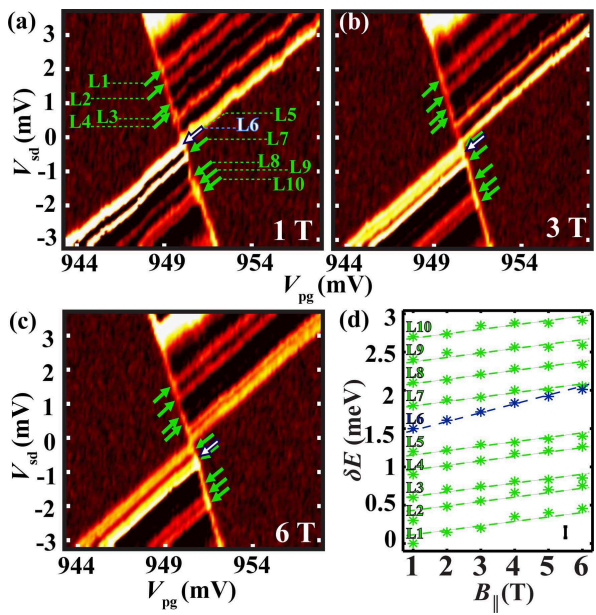}
\caption{ \label{fig4} (Color online) (a--c) Stability diagrams of Structure A in magnetic fields ranging from~1 to 6~T . The spin excited state [L6 in panel (a)] is marked with a white and blue arrow, and the DOS lines are labeled with green arrows. (d) Shift of the conductance lines in magnetic field. Data for each line is offset in energy for clarity. The dashed lines show a slope $2\mu_B$ for L6 and $\mu_B$ for the other data.}% end capti
\end{figure}

The stability diagrams shown in Fig.~\ref{fig3}(a,d) do not provide
enough information to distinguish whether a specific conductance
feature is due to the reservoir DOS or an internal excited state.
However, the lead-gate-compensated traces shown in
Fig.~\ref{fig3}(b,c,e,f) reveal the origin of the conductance peaks.
Most of the peaks move with the lead gate voltage, and hence
correspond to the reservoir DOS. However, Fig.~\ref{fig3}(b,c) shows
a feature $0.4$~meV above the ground state for all $V_\textrm{lg}$.
As justified by measurements carried out in various magnetic fields
(see Fig.~\ref{fig4}), this line corresponds to the spin excited
state of the neutral donor orbital ground state. Similarly, the
vertical lines in Fig.~\ref{fig3}(e,f) are attributed to excited
states of the dot.

For Structure A, the slope of the DOS lines yields the response of
the Fermi level $E_F$ to the lead gate voltage to be $\Delta
(E_F-E_c)/\Delta V_\textrm{lg}= (11.3 \pm 1.5)$~meV/V, where $E_c$
is the conduction band minimum. We have modeled this response using
technology computer aided design~\cite{tcad} taking into account the
three-dimensional geometry of the system. By matching the measured
response with the model, we obtain an estimate of the absolute value
of the Fermi level to be $E_F-E_c= (25 \pm 5)$~meV.

The average energy spacing between the DOS peaks is obtained from
Fig.~\ref{fig3} to be $(0.33\pm 0.03)$~meV for Structures A and B.
We model this by calculating the single-particle energy spectrum for
a triangular well $V_z(z)=eF|z|/\Theta(-z)$ in the $z$ direction
($\Theta$ is the Heaviside step function), an infinite $W=70$~nm box
potential in the $y$ direction, and a semi-infinite box potential in
the $x$ direction. Here, $F$ is the electric field in the vicinity
of the Si--SiO$_2$ interface due to the lead gate. The
one-dimensional DOS in the $x$ direction, for which the continuum
approximation is valid, is given by
$D_\textrm{1d}(E)=\sqrt{2m_x}/(\pi\hbar\sqrt{E})$, where $m_x$ is
the effective mass of the electron in the $x$ direction. Thus we
obtain for the total density of states
\begin{equation}\label{eq1}
D(E)=\frac{\sqrt{2}}{\pi\hbar}\!\sum_{n_y,n_z,n_v}\!\!\!\frac{\sqrt{m_{x,n_v}}\Theta[{E-E_{y,z}(n_y,n_z,n_v)}]}{\sqrt{E-E_{y,z}(n_y,n_z,n_v)}},
\end{equation}
where $n_y$, $n_z$, and $n_v$ are the indices for the energy levels
in the $y$, $z$, and valley degrees of freedom, respectively. The
six-fold valley degeneracy of bulk silicon is lifted due to
different effective masses $m_t=0.190\times m_0$ and
$m_l=0.916\times m_0$ for the transverse and longitudinal
directions, respectively~\cite{AndoRMP}. Here, $m_0$ is the electron
rest mass. Thus the effective mass in the $x$ direction depends on
the valley degree of freedom. According to Eq.~\eqref{eq1}, a peak
in DOS is observed at energies matching the single-particle energy
levels in the two-dimensional $yz$ potential, i.e., for
$E=E_{y,z}(n_y,n_z,n_v)$.

The peak spacing resulting from the confinement in the $y$ direction
is given by $\pi^2\hbar^2(n_y+1/2)/(m_yW^2)$. This yields the
minumum peak spacing of 1.2~meV for the transverse and 0.25~meV for
the longitudal effective masses. Each excited state in the valley
and $z$ degree of freedom superimposes an additional series of
peaks, and hence the resulting DOS structure can be rather
complicated. Figure~\ref{fig3}(g) shows the DOS in the vicinity of
the Fermi level obtained above for Structure A. The electric field
of is chosen to be $F = 2$~MV/m to match the peak separation
observed in the experiments. This field is approximately an order of
magnitude lower than expected due to the crude model used to
calculate the peak spacing. This simple model does not include the
true details of the trapping potential including disorder or the
electric field induced by the accumulation layer, which calls for
deeper theoretical analysis. Furthermore, we have not included the
possible splitting of all valley degeneracies which would further
reduce the average peak spacing and consequently, increase the
required electric field.

Figure~\ref{fig1}(c) shows a schematic diagram of how the DOS peaks
behave in magnetic field $\mathbf{B}=B\hat{\mathbf{y}}$. The Zeeman
effect shifts the kinetic energy at the Fermi level ($E_F-E_c$) for
the spin-down electrons up by $g_L\mu_B B/2$, and vice versa for
spin-up electrons. Here, $g_L$ is the electron $g$ factor in the
reservoirs and $\mu_B=e\hbar/(2m_0)$. Thus the effect of this term
is to shift and split the DOS peaks. However, if the electron $g$
factor for the donor or dot, $g_D$, is almost the same as $g_L$,
this compensates for the splitting and results only in a downwards
shift of the DOS peaks by $g_L\mu_B B/2$ with respect to the
transport window. In our case, the DOS peaks simply shift and do not
split in magnetic field, as demonstrated in Fig.~\ref{fig4}, which
implies $g_D\approx g_L\approx 2$ consistent with previous
measurements~\cite{Willems2008}. In contrast, Zeeman splitting of
DOS peaks has been observed in tunneling through a Si shallow donor
in a GaAs/AlAs/GaAs junction~\cite{Jouault2009}.

Figure~\ref{fig4}(d) shows the energy shift $\delta E$ of the
conductance peaks with respect to the transport window as a function
of magnetic field. We observe that within the estimated error, the
shift of the DOS peaks and the spin excited state is in agreement
with $\delta E = \mu_B B$ and $\delta E = 2\mu_B B$, respectively.
Thus our observations can be explained by the Zeeman effect alone.

In conclusion, we have reported the first studies of reservoir DOS
in nanostructures through gate control of the reservoir electron
density. Magnetic field spectroscopy of the DOS peaks revealed
behavior consistent with energy shifts due to the Zeeman effect
alone. Both the single donor and the quantum dot devices studied
exhibited similar conductance patterns arising from the reservoir
DOS. Our findings not only confirm the interpretation that
additional conductance lines observed previously in transport
through single donors are due to reservoir
states~\cite{SellierPRL06,LansbergenNatPhys08,Tan2009}, but also
demonstrate that these states can be controlled and studied in a
quantitative manner.

The authors thank F.\ Hudson, D.\ Barber, and R.\ P.\ Starrett for
technical support, E.\ Gauja, A.\ Cimmino, and R.\ Szymanski for
assistance in nanofabrication, and M. Eriksson, J.\ P.\ Pekola, and
V.\ Pietil\"a for insightful discussions. M.\ M. and J.-M.\ P.\
acknowledge Academy of Finland, Emil Aaltonen Foundation, and
Finnish Cultural Foundation for financial support. This work is
supported by the Australian Research Council, the Australian
Government, the U.S. National Security Agency (NSA), and the U.S.
Army Research Office (ARO) (under Contract No. W911NF-08-1-0527).

\bibliography{manu}

\begin{thebibliography}{26}
\bibitem[{FN1()}]{FN1x}
\emph{\bibinfo{title}{\rm \textsc{A}n increment in the lead gate voltage also
  results in a slight increase of the transparency of the tunnel barriers.
  \textsc{I}n \textsc{S}tructure \textsc{B}, this can be compensated by
  decreasing the voltage on the barrier gates.}}
\end{thebibliography}
\end{document}